\documentstyle[multicol,prl,epsfig,aps]{revtex}
\begin{document}
\draft
\title{First time determination of the microscopic structure of a stripe
phase:\\ Low temperature NMR in La$_{2}$NiO$_{4.17}$}

\author{ I.M. Abu-Shiekah, O. Bakharev, and H.B. Brom}
\address{Kamerlingh Onnes Laboratory, Leiden University, P.O.Box 9504, 2300 RA\\
Leiden, The Netherlands}
\author{J. Zaanen}
\address{Instituut Lorentz for Theoretical Physics, Leiden\\
University, P.O. Box 9506, 2300 RA Leiden, The Netherlands}
\date{Phys. Rev. Lett. {\bf 87}, 237201, 3 December 2001}
 \maketitle

\begin{abstract}
The experimental observations of stripes in superconducting cuprates
and insulating nickelates clearly show the modulation in charge and
spin density. However, these have proven to be rather insensitive to
the harmonic structure and (site  or bond) ordering. Using
$^{139}$La NMR in La$_{2}$NiO$_{4+\delta}$ with $\delta=0.17$, we
show that in the $1/3$ hole doped nickelate below the freezing
temperature the stripes are strongly solitonic and site ordered with
Ni$^{3+}$ ions carrying $S=1/2$ in the domain walls and Ni$^{2+}$
ions with $S=1$ in the domains.
\end{abstract}


\begin{multicols}{2}
\settowidth{\columnwidth}{aaaaaaaaaaaaaaaaaaaaaaaaaaaaaaaaaaaaaaaaaaaaaaaaa}

Stripe phases have by now been observed in a variety of doped Mott-insulators,
like cuprates, nickelates and manganites\cite{Emery99}. Nevertheless there is
still a remarkable lack of knowledge on the details of the charge- and spin
distributions in these novel electronic phases. At the same time there is a
growing body of theoretical literature dealing with the microscopic origin of
stripe formation, predicting stripes starting from rather different physical
perspectives\cite{Emery99,Zaanen98,White98,Vojta00,Rome01}. Given its potential
connections for instance to the mechanism of superconductivity, it is a matter of
high urgency to find out how the stripes look like in detail. In this regard the
various theories lead to quite different predictions. Because the spin- and
charge distributions are inhomogeneous, NMR with its microscopic sensitivity
could in principle yield important information. However, in cuprates attempts in
this direction have been frustrated due to the anomalous, glassy ordering
dynamics obscuring the static signal. Neutron scattering shows that both in
cuprates and nickelates the correlation length of the stripe order is relatively
small.  This disorderly nature comes to play in NMR \cite{Teitelbaum00,Hunt00} in
the form of a peculiar loss of signal intensity
(wipe-out)\cite{Teitelbaum00,Hunt00,Hunt99,AbuShiekah99,Curro00,Suh00} explained
by a spread in very short spin-dephasing times
$T_2$\cite{Teitelbaum00,Curro00,Suh00}. Even at the lowest temperatures the
signal recovery is only partial indicating that this dynamics is still at work at
temperatures as low as 0.3 K in the cuprates, prohibiting attempts to deduce
information about the static order from the NMR data \cite{Hunt00}. We
demonstrate here that in a nickelate stripe system the nature of the stripe order
(established by neutron scattering) can be deduced in detail from NMR, despite
the strong similarities with the cuprate NMR at higher temperatures. We find the
stripe structure to be strongly solitonic, with sharply defined charge stripes
with a width which is not exceeding the lattice constant by much. Surprisingly,
they look quite like the site centered stripes predicted by early mean-field
calculations for the nickelate system\cite{Zaanen94}.

Before analyzing the low temperature line shape, which is the central issue of
this Letter, we first introduce the main features of the NMR line shape, derive
the value of the hyperfine coupling and explain the partial recovery of the
signal intensity from the relaxation data.

The NMR measurements were performed on the same La$_{2}$NiO$_{4.17}$ crystals as
measured before\cite{AbuShiekah99,Bernal97}, by sweeping the external field $B$
at various frequencies. The crystal symmetry is almost tetragonal\cite{Mehta94},
which makes the $^{139}$La-spectra ($I=7/2$) strongly dependent on the direction
of the magnetic field with respect to the crystallographic axes.

\begin{figure}[htb]
\begin{center}
\leavevmode \epsfig{figure=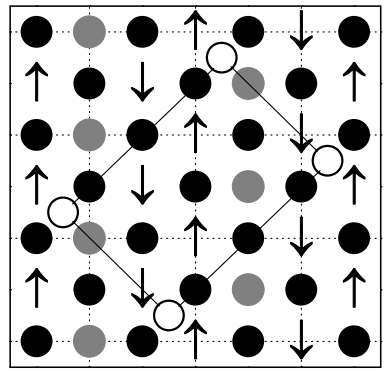,width=5cm,angle=0}
\end{center}
\caption{Ordering of excess oxygen, holes and spins. Open circles refer to the
ordered excess oxygen sites at $z=0.217c$, which introduce holes assumed to be
Ni$^{3+}$ with $S=1/2$ (gray circles, site-order scenario). The arrows refer to
the Ni$^{2+}$ spins with $S=1$. Black circles indicate the $^{139}$La sites at
$z=0.14c$. The dotted lines are the in-plane boundaries for the unit cell before
doping. The drawn tilted rectangle is the enlarged unit cell after doping.}
\label{excessoxygen}
\end{figure}

In La$_{2}$NiO$_{4.18}$, with hole doping very close to
La$_{2}$NiO$_{4.17}$, details about the excess oxygen atoms
positions are available\cite{Mehta94}. The interstitial oxygen
sites are at (0.183a, 0.183b, 0.217c) or equivalent positions. A
scenario for the excess oxygen ordering is shown in
Fig.\ref{excessoxygen}. There is one excess oxygen atom for each
six unit cells. The number of $^{139}$La sites, which are far from
the excess oxygen atoms, is twice as large as the number closest
to the excess oxygen atoms. For La$_{2}$NiO$_{4.17}$, using simple
point charge calculations, we expect the interstitial oxygen to
change the electrical field gradient (EFG) around the La-sites and
to split the $^{139}$La NMR line into two: an A-line due to
A-sites with the main component of the EFG ($V_{zz}$) along the
$c$-axis and a B-line due to two times less abundant B-sites with
the crystal field gradient in the ($ab$)-plane. This prediction is
in agreement with the experimental intensity ratio of line A to B
\cite{AbuShiekah99,AbuShiekahthesis01}.

\begin{figure}[htb]
\begin{center}
\leavevmode \epsfig{figure=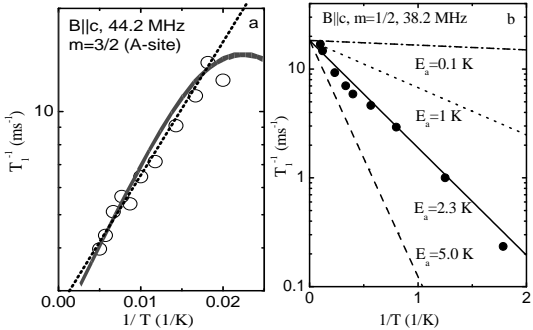,width=7cm,angle=0}
\end{center}
\caption{\sl (a) $T_1^{-1}$ for line A (see text) for $m$=3/2 at
44.196~MHz, obtained from stretched exponential fits to the
recoveries. The dotted line is a fit with $E_{0}=90$~K. Drawn line
is a fit using a half Gaussian distribution in activation energy.
(b) $T_1^{-1}$ versus $1/T$ at low temperatures. An activation
energy of 2.3~K fits the data well. Sites with lower activation
energies will not be observable ($m=1/2$ refers to the $+1/2,-1/2$
and $m=3/2$ to the $3/2,1/2$ transition.}\label{T1high}\label{t1low}
\end{figure}

Above 50~K, the $^{139}$La relaxation rates can be fitted by an
Arrhenius expression $T_{1}^{-1}\propto \exp (E_{0}/T)$, see Fig.
\ref{T1high}a \cite{AbuShiekah99}. Fits allow for a (full or half
gaussian) distribution in activation energies around $E_{0}=90$~K
with a width of $\Delta=25$~K. Above the magnetic freezing
temperature the rates for line A are at least twice the rates of
line B, and have almost the same temperature dependence
\cite{AbuShiekah99}. To extract the amplitude of the fluctuating
fields and its correlation time from the relaxation data, the value
of the hyperfine constant is needed. This constant can be determined
in several ways. In doped La$_{2}$CuO$_{4}$ the static internal
magnetic field amounts to about 0.1~T \cite{Teitelbaum00}, which is
similar to the value found in the undoped case \cite{MacLaughlin94}.
Because the hyperfine field at the lanthanum site in the nickelates
is calculated to be about 18 times larger than in the cuprates
\cite{Takahashi91}, its value will be around 1.8~T. Yoshinari
$\it{et~al.}$ \cite{Yoshinari99} reported a value of about $2.3\pm
0.5$~T/$\mu_{B}$ for the hyperfine coupling in Sr-doped nickelates
with an ordered moment of $\approx 0.7~\mu_{B}$. The so obtained
internal field of 1.6~T agrees well with the internal field of about
1.78~T reported by Wada {\it et al.} \cite{Wadat93} for
La$_{2}$NiO$_{4.10}$. Also by comparing our susceptibility and
linewidth data at high temperature we can estimate the hyperfine
field to be about 2.0~T, consistent with the earlier quoted values
\cite{Bernal97}. Using a hyperfine field of 1.8~T, the fit to the
relaxation data gives a value of 2600~s$^{-1}$ for $\gamma
_{n}^{2}h_{0}^{2}\tau _{\infty }$, and hence $\tau _{\infty
}=1.8\times 10^{-11}$~s. $T_2^{-1}$ has almost the same $T$
dependence as $T_{1}^{-1}$. It deviates only for $T> 200$~K.

The relaxation data below 20~K can not be fitted by the same parameters, see Fig.
\ref{t1low}b. The rates can be reproduced by $T_{1}^{-1}\propto \gamma^2 h_0^2
\exp (-E_{0}/T)/(\omega_0^2 \tau _{\infty })$, with $E_{0}= 2.3 \pm 0.2$~K and
$h_0$ the strength of the fluctuating field. Measurements of the rates were
performed for different satellites at 38.2~MHz. Using the same methods as
described in Ref.\onlinecite{Teitelbaum00}, a magnetic origin for the fluctuating
fields is found at 160~K, 90~K and 1.6~K.

The wipe-out of the $^{139}$La NMR signal intensity and its partial
reappearance below $\sim$ 11~K \cite{AbuShiekah99} can be explained
in the same terms as used before in the cuprates \cite{Teitelbaum00}
indicating a distribution in spin dephasing times, hence activation
energies\cite{AbuShiekahthesis01}. At low temperatures only nuclei
relaxing with activation energies $>E_{0}$ lead to the measured
signal, while the others will have dephasing rates higher than
$10^{4}$~Hz down to very low temperatures. Even at $T \approx $
0.5~K slow spin fluctuations apparently prevent the full recovery of
the signal. The intensity recoveries can be fitted with
$\Delta=2.8$~K and $E_{0}\simeq 0$~K\cite{AbuShiekahthesis01}. These
results are consistent with the relaxation data because only nuclei
with the higher activation energies will be visible.

What are the implications of these observations? In the renormalized classical
limit below charge ordering the spin correlation rate $\tau(T)$  will be
proportional to the spin correlation length $\xi (T)$ divided by the spin wave
velocity $c$: $T_{1}^{-1} \sim (\xi/c)(T/2\pi\rho)^2(1+T/2\pi\rho)^{-2}$, with
$\rho$ the spin stiffness\cite{Teitelbaum00,Hunt00}. In La$_{2}$NiO$_{4.17}$
according to diffraction studies the stripes have a finite and temperature
independent correlation length\cite{Du00,Lee01}. If this has a dynamic origin, it
will be consistent with a distribution in activation energies or $\tau$'s.
Possible sources for such a distribution in the spin dynamics are dislocations
\cite{Zaanen00,Zachar00} in the domain pattern or elastic deformations
\cite{Zachar00}.

We now come to the central part of this paper. We will show that although the
interstitial oxygens have a direct influence on the La signal via the electric
field gradients, it is still possible using $^{139}$La NMR to probe the charge
and spin ordering in the NiO$_2$ plane.

\begin{figure}[htb]
\begin{center}
 \leavevmode \epsfig{figure=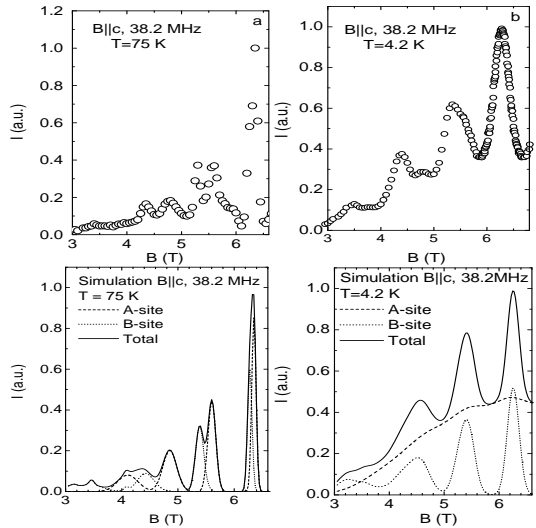,width=7cm,angle=0}
\end{center}
\caption{High (a) and low (b) $T$ spectra for $B \parallel c$ and
their simulations (lower panels). The simulations give precise
values for the EFG parameters but are not very sensitive for the
strength of the internal field (at low $T$). Parameters are
discussed in the text. } \label{lowfit}
\end{figure}

\begin{figure}[htb]
\begin{center}
 \leavevmode \epsfig{figure=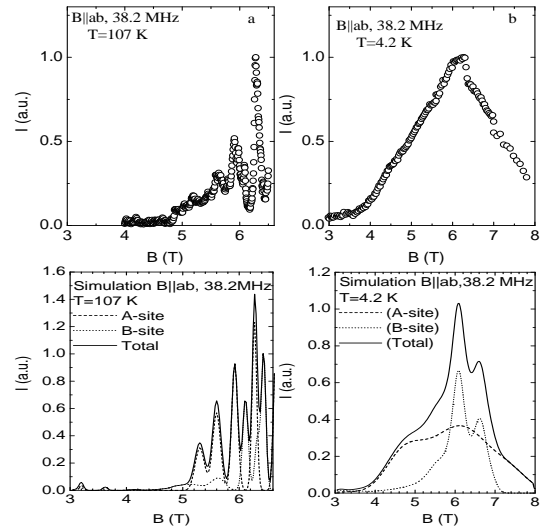,width=7cm,angle=0}
\end{center}
\caption{High (a) and low (b) $T$ spectra for $B \parallel ab$ and their
simulations (lower panels). Spectra are simulated with the same parameters as in
Fig.\ref{lowfit}. The internal field leads to a splitting of the lines and hence
can be determined precisely.}\label{lowfitab}
\end{figure}

 Below 15~K, when the signal starts to
regain some of its intensity, the NMR lines are largely broadened. In Figs.
\ref{lowfit},\ref{lowfitab} the spectra are shown for $B
\parallel c$ and $B \perp c$ . Simulations of the spectra taking into account the quadrupolar
and Zeeman contributions \cite{Slichter91} are shown in the lower
panels. The spectra can be decomposed into 2 La sites, one with
resolved satellites and the other with overlapping lines due to
much stronger magnetic broadening. Magnetic freezing or ordering
is also visible in the bulk magnetic susceptibility, which starts
to deviate from the Curie law and peaks at $T=17\pm
3$~K\cite{AbuShiekah99,Bernal97,Odier99}. The peak in $ - d\chi
/dT$ indicates a series of cusps at different spin freezing
temperatures close to 17~K and can be due to spin frustration or
spin clusters\cite{Odier99}.

We first show that contrary to the observation in the cuprates \cite{Hunt00}, the
low temperature line pattern in the nickelate is not due to motional narrowing.
In case of two lines separated by $\delta \omega$ the condition for motional
narrowing \cite{Teitelbaum00,Slichter91} is that the correlation time $\tau$ for
the (spin) fluctuations is much less than $1/\delta \omega$. The splitting
$\delta \omega$ is given by $\gamma_n$ times the value of the internal field
$h_0$: $\delta \omega = \gamma_n<h_0>$. The value of $\tau$ is coupled to the
relaxation rate via $\tau \sim \gamma_n^2 <h_0^2> T_1/\omega^2$. Hence the
condition can be rewritten as $R_m =\gamma_n^3 <h_0^2>^{3/2}T_1/\omega^2 \ll 1$.
Using typical values for $h_0 = 1.5$~T, $\omega = 2\pi \times 40$~MHz and $T_1 =
50$ $\mu$s, $R_m \sim 10^2$.

What is the information we might extract from the lineshape data?
The high $T$ data will be sensitive for the EFG parameters and the
low $T$ data in addition for the internal or local field. Since
the internal field lies in the $ab$-plane, the spectra for $B
\parallel c$ will be rather insensitive to the local field. For $B
\parallel ab$ the antiferromagnetic alignment of the electron
spins we expect to lead to a splitting of the lines. The
simulations of the spectra are shown in the lower panels. The
principal axes of the EFG with respect to the crystal axes are
described by Eulerian angles ($\alpha, \beta ,\gamma $)
\cite{Goldstein50} and the external magnetic field is described by
polar angles ($\theta ,\varphi $) with respect to crystallographic
axes.

 For $B
\parallel c$ ($\theta=0$) the high $T$ data (e.g. at 75~K) at $\nu =$~38.2 MHz
are well reproduced by a quadrupolar splitting $\nu_Q = eQ V_{zz}/4
\hbar I(2I-1)$ (with $Q$ the electric quadrupole moment of the
$^{139}$La-nucleus) of 4.5 MHz. The anisotropy parameter
$\eta=(V_{xx}-V_{yy})/V_{zz} = 0$. The width of the satellites in
the spectra can be simulated by introducing a spread in the EFG
parameter $\Delta \nu_Q = 0.5$~MHz, while the width of the main line
requires a (dipolar) field distribution of 0.05~T at the La(A)-site.
For the La(B)-site these values are $\nu_Q$=8.5~MHz, $\eta=0.75$,
$\Delta \nu_Q =1.0$~MHz and $\Delta_B=0.05$~T. The other parameters
are for the A-site: $\alpha =0,$ $\beta =0,$ $\gamma =0$; for the
B-site: $\alpha =0,$ $\beta =\pi /2,$ $\gamma =-\pi /6$. According
to the simulation of the low temperature data the main effect of the
magnetic freezing is the increase of the magnetic broadening
($\Delta_B)$ to 0.5, resp. 0.15~T. The spectra shown use internal
fields of 1.5~T at the A-site and 0.5~T at the B-site. These
internal field values follow from the simulations for $B\perp c$,
where the values are much more restricted. For the $B\perp c$
orientation simulation of the 107~K data with the same parameter set
as for $B \parallel c$ reproduce the data well ($\varphi =\pi /6$).
For the 4.2~K data the antiparallel alignment of the internal field
leads to a splitting of the lines and is described by polar angles
($\theta =\pi /2,\varphi =\pi /3\pm \pi /2$) with respect to
crystallographic axes.

To explain these remarkable differences, let us assume that we deal with site
centered stripes. These stripes naturally lead to two kinds of Ni-ions: Ni$^{2+}$
(A-sites) and Ni$^{3+}$ (B-sites) with two kinds of spins $S=1$ and $S=1/2$
respectively. As a consequence the La sites will experience different hyperfine
fields. The internal dipolar magnetic field arising from the Ni-spins at the
La-sites will be of the order of 0.2~T. However, the La nuclei just above and
below Ni$^{2+}$ in addition will have an exchange coupling via the oxygens. This
hyperfine field due to the overlap of the Ni 3d$_{z^{2}}$ and La 6s orbitals
through the 2p$_{z}$ orbital of apical oxygen, is about 1.8~T in the undoped
samples and is almost doping independent\cite{Furukawa94}. Due to the different
occupation of the 3d$_{z^{2}}$ orbital of Ni$^{3+}$ the exchange coupling between
the $S=1/2$ Ni$^{3+}$ spins and the La sites will be
weaker\cite{Pellegrin96,Anisimov99}. The ratio between $^{139}$La sites that feel
the exchange field of Ni$^{3+}$ and Ni$^{2+}$ will be close to 1:2. This
difference in intensity ratio and hyperfine field are indeed the main
characteristics of the line shape and hence are well accounted for by this
scenario. Can bond centered stripes explain the observations as well? The
hyperfine fields will have the same maximum, but the distribution will be
different. The line shapes for $B\perp c$ puts a limit to the field on the B
sites of at most 0.3 T, which rules out this possibility.

The experiments show that apart from the internal field we need to introduce an
appreciable magnetic broadening. Part of the broadening might be due to canting
of the spins in the ordered phase in the NiO$_2$-plane away from the charge and
spin stripe direction\cite{Lee01}, which effect we have not included into this
calculation. Another reason for the extra broadening might be found in the finite
size of the correlated magnetic regions.

Summarizing, in La$_{2}$NiO$_{4.17}$ interstitial oxygens
determine the line profile above the wipe-out temperature. The NMR
intensity loss above the spin freezing or ordering temperature
around 20~K is linked to a spread in spin-dephasing as in the
cuprates. From the angular and temperature dependence of the La
line profiles, we show that the distribution of the internal
fields is in agreement with two kinds of Ni-sites with different
ionicity and hence different hyperfine interaction with the
visible La sites. Site centered stripes of the kind predicted by
mean-field theory\cite{Zaanen94} fit the low temperature data of
the visible La-nuclei remarkably well.

We like to acknowledge A.A. Menovsky, A.A. Nugroho and Y. Mukosvkii
for providing the nickelate samples, O.O. Bernal and P.M. Paulus for
useful discussions, and O. Berfelo for performing the TGA. This work
was supported by FOM-NWO.

\end{multicols}

\end{document}